\documentclass{elsart}
\usepackage{epsfig}
\usepackage{amssymb}

\begin{document}

\begin{frontmatter}

\title{Dependence of volume FEL (VFEL) threshold conditions on undulator parameters}

\author{V. Baryshevsky, K. Batrakov}
\address{Research Institute of Nuclear Problem, Belarussian State University,\\
11 Bobruyskaya Str., 220050, Minsk, Belarus}

\begin{abstract}
Volume free electron laser uses volume distributed feedback (VDFB),
which can essentially reduce the threshold current of generation and provides
the possibility of smooth frequency tuning.
An undulator VFEL is considered.
\end{abstract}


\end{frontmatter}

\section{Introduction}
\qquad The advantages of distributed feedback (DFB) is well known
in laser physics.
In conventional lasers DFB is formed in such a way that
diffracted and transmitted waves propagate
along a line in opposite directions.
The distinctive feature of volume free electron lasers (VFEL) is using of
non-one-dimensional (volume) distributed feedback at which the diffracted wave 
propagates at the angle $\ne \pi$ to the transmitted wave.
Firstly the idea to use
volume distributed feedback (VDFB) for X-ray generation was proposed 
in  \cite{bar1}.
VDFB causes sharp increase of amplification 
and instability increment of electron beam
density is proportional to $\sim \rho^{1/(3+s)}$, 
where $\rho$ is the electron beam
density and $s$ is the number of surplus waves 
appearing due to diffraction (for example,
in case of two-wave Bragg diffraction $s=1$, 
for three-wave diffraction $s=2$ and so on).
This dependence essentially differs from increment for conventional
FEL in Compton regime, which is proportional to $\sim \rho^{1/3}$ \cite{marsh}.
Now the investigation of 4-wave diffraction in system with two-dimentional DFB 
is also started by \cite{Ginzb}.

\qquad Sharp increase of amplification caused by VDFB yields to noticeable reduction 
of threshold currents necessary for lasing start. 
This fact is particularly important 
for lasing in submillimetre and visible ranges and for shorter wavelengths. 
Explicit expressions VFEL
threshold currents were obtained in \cite{ba2}.
In present work the dependence of VFEL starting current on 
undulator parameters is considered.

\section{Generation and amplification laws of undulator VFEL}
\begin{figure}[h]
\epsfxsize = 14 cm
\centerline{\epsfbox{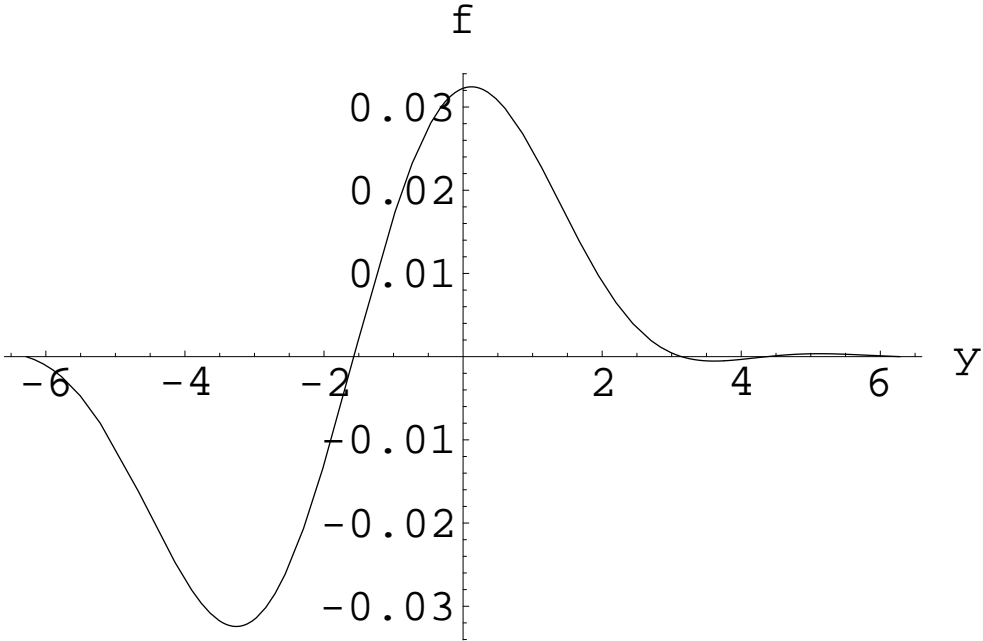}}
\caption{Dependence of induced radiation on detuning factor $y$ in the
condition of two-wave diffraction.}
\label{curve}
\end{figure}

\qquad It is well known that to find amplification and starting current one 
should study the dispersion law of radiating system
The set of equations describing interaction of relativistic electron beam,
which propagates in spatially periodic structure in undulator is \cite{ba2}:
\begin{eqnarray}
DE-\omega ^{2}\chi _{1}E_{1}-\omega ^{2}\chi _{2}E_{2}-...=0 \nonumber \\
-\omega ^{2}\chi _{-1}E+D_{1}E_{1}-\omega ^{2}\chi _{2-1}E_{2}-...=0  \label{system} \\
-\omega ^{2}\chi _{-2}E-\omega ^{2}\chi _{1-2}E_{1}+D_{2}E_{2}=0-..., \nonumber \\
...\nonumber
\end{eqnarray}
where $D_{\alpha }=k_{\alpha }^{2}c^{2}-\omega ^{2}\varepsilon +\chi
_{\alpha }^{(b)}$ , $\vec{k}_{\alpha }=\vec{k}+\vec{\tau}_{\alpha }$ are the
wave vectors of photons diffracted by the crystal planes with corresponding
reciprocal vectors $\vec{\tau}_{\alpha }$, $\varepsilon_{0} =1+\chi _{0}$ , $\chi
_{\alpha }$ are the dielectric constants of a periodic structure. These constants
can be obtained from the following representation of dielectric permiability of
periodic structure:
\begin{eqnarray*}
\varepsilon (\vec r,\omega )=1+\sum_{\{\tau \}}\exp (i\vec \tau \vec r%
)\chi _{\tau }(\omega).
\end{eqnarray*}
$\chi _{\alpha }^{(b)}$ is the part of dielectric susceptibility appearing from
the interaction of an electron beam, propagating in undulator, with radiation:
\begin{eqnarray}
\chi _{\alpha }^{(b)} &=&\frac{\pi \Theta _{s}^{2}c^2}{\gamma _{z}^{2}\gamma I_{A}}
\frac{j_{0}}{(\omega -
(\mathbf{k}+\mathbf{k}_w)\mathbf{u}_w)^{2}} \nonumber   \\
&& for\ \ \  the\ \ \ "cold"\ \ \  beam\ \ \  limit\ \ \  and \label{beam}  \\
\chi _{\alpha }^{(b)} &=&-i\sqrt{\pi }\frac{\pi \Theta _{s}^{2}c^2}{\gamma _{z}^{2}\gamma I_{A}}
\frac{j_{0}}
{\sigma _{\alpha }^{2}}x_{\alpha }^{t}\exp [-(x_{\alpha }^{t})^{2}]
\nonumber \\
&& for\ \ \  the \ \ \ "hot" \ \ \  beam \ \ \  limit,  \nonumber
\end{eqnarray}
$\Theta _{s}=eH_{w}/(mc^{2}\gamma k_{w})$, $\gamma _{z}^{-2}=\gamma
^{-2}+\Theta _{s}^{2}$, $k_{w}=2\pi/\lambda_w$, $\lambda_w$ is undulator period,$H_w$ is undulator field,
$x_{\alpha}^{t}=(\omega -
(\mathbf{k}+\mathbf{k}_w)\mathbf{u}_w)/\sqrt{2}\sigma _{\alpha }$, $\sigma
_{\alpha }^{2}=(k_{\alpha 1}^{2}\Psi _{1}^{2}+k_{\alpha 2}^{2}\Psi
_{2}^{2}+k_{\alpha 3}^{2}\Psi _{3}^{2})u^{2}$ and $\vec{\Psi}=\Delta \vec{u}/u$
is the velocity spread. If the inequality $x_{\alpha}^{t} \gg 1$ is fulfilled,
all the electrons interact with electromagnetic wave and the "cold" limit is realized.
In the opposite case ("hot" limit)  $x_{\alpha}^{t} < 1$ only small part of electron beam
interacts with electromagnetic wave.
Setting the determinant of linear system (\ref{system}) equal to zero
one can obtain the dispersion equation for the system
"electromagnetic wave + undulator + electron beam + periodic structure".
In case of two-wave dynamical diffraction this equation has the following form:
\begin{eqnarray}
DD_1-\omega^4\chi _1\chi _{-1}=0 \label{two}
\end{eqnarray}
For the system with finite interaction length the solution of boundary problem
can be presented as a sum:
\begin{eqnarray}
\mathbf{E}+\mathbf{E}_{1}=
\sum_{i}c_{i}\exp \{i\mathbf{k}_{i}\mathbf{r}\}(\mathbf{e}+\mathbf{e}
_{1}s_{1}^{(i)}\exp \{i\mathbf{\tau r}\}), \label{bound2}
\end{eqnarray}
here $s_{1}^{(i)}=\frac{k_i^2 c^2-\omega^2\varepsilon _0}{\omega^2\chi _{1}}$
are the coupling coefficients between the diffracted and transmitted waves
($E^{(1)}=s_{1}E$) and $\vec{k}_i$ are the solutions of dispersion equation (\ref{two}). 
To determine
coefficients $c_i$ it is necessary to write the boundary conditions on 
the system ends
$z=0$ and $z=L$. 
For Bragg geometry, when transmitted and diffracted waves has
the opposite signs of wave vector projections on the axis $z$, these conditions are as follows:
\begin{eqnarray}
\sum_{i}c_{i}=a\ \ \  \sum_{i}\frac{c_{i}}{\delta _{i}}=0\  \  \
\sum_{i}\frac{c_{i}}{\delta _{i}^{2}}=0 \label{conditions} 
 \ \ \  \sum_{i}s_{1}^{(i)}c_{i}\exp \{ik_{iz}L\} =b 
\end{eqnarray}
{In (\ref{conditions}) the wave vector is represented as:}
$\mathbf{k}=\mathbf{k}_0+\frac{\omega}{c}\delta$, where $\mathbf{k}_0$ satisfies
undulator synchronism condition. The boundary conditions (\ref{conditions}) are written
for the "cold" electron beam. In this case the dispersion equation has four roots 
($\delta_i$, $i=1\div 4$).
The first and the fourth conditions in (\ref{conditions}) correspond to continuity of
transmitted wave at $z=0$ and diffracted wave at boundary $z=L$ (it is supposed that the wave
with wave vector $\vec{k}$ and amplitude $a$ is falling on boundary $z=0$ and
the wave with wave vector $\vec{k}_1$ and amplitude $b$ is falling on boundary $z=L$).
The second and the third conditions in (\ref{conditions}) accord with the
requirement that the electron beam
is unpertubed before entering the interaction region.
The part of electron beam energy converting to radiation can be expressed as:
\begin{eqnarray}
I\sim \gamma _0 |E(z=L)|^2+|\gamma _1| |E_1(z=0)|^2= \label{energy} \\
(\gamma _0 |a|^2+|\gamma _1| |b|^2)\left(\frac{\gamma _0 c}{\vec{n}\vec{u}}\right)^3
\frac{16 \pi^2 n^2}{-\beta (k|\chi _1|L_*)^2kL_*(\Gamma _{start}-\Gamma)},
\nonumber
\end{eqnarray}
where $L$ is the length of interaction in undulator,
\begin{eqnarray*}
\Gamma _{start}=\left(\frac{\gamma _0 c}{\vec{n}\vec{u}}
\right)^3\frac{16 \pi^2 n^2}{-\beta (k|\chi _1|L_*)^2}-
\chi"\left(1-\beta \pm \frac{r"\sqrt{-\beta}}{|\chi _1|\chi"}\right) \\
\Gamma =\frac{\pi^2 n^2}{4}\frac{\pi  \Theta _{s}^{2}c^{2}j_0}
{\gamma _{z}^{2}\gamma  I_{A} \omega^2} k^2 L_*^2 q^2 f(y),\\
f(y)=\sin y \frac{(2y+\pi n)\sin y-y(y+\pi n)\cos y}{y^3(y+\pi n)^3}
\end{eqnarray*}
is the function of generation dependence on detuning from synhronism condition, 
$y=(\omega -\mathbf{k}\mathbf{v}_w-\Omega )L/(2u_z)$ is detuning factor,
$\beta=\gamma_0/\gamma_1$ is diffraction asymmetry factor, $\gamma_0$, $\gamma_1$  are diffraction cosines,
$\chi" =Im\ \chi_0$. The function $f(y)$ is presented in fig.\ref{curve}
One can see from fig.\ref{curve} that dependence on detuning factor $y$ is not asymmetric.
This distinguishes lasing in the range of roots degeneration from generation process
in conventional undulator FEL. The latter has the 
following dependence on detuning factor \cite{marsh}: 
$$
g(y)=\frac{\sin y}{y}\frac{y\cos y-\sin y}{y^2}.
$$
This difference ensues from interference of contribution to radiation of two diffraction 
roots.
From (\ref{energy}) follows that:
1. the starting current in case of two-wave diffraction is proportional to
$j_{start}\sim (kL)^{-1}(k\chi _1L)^{-2} $; \\
2. non-one dimensional VDFB provides the possibility to decrease the starting current of 
generation by varying of the angles between the waves. 
The dependence of $\Gamma _{start}(\beta)/\Gamma _{start}(\beta=0)$
on asymmetry factor $\beta$ is presented in fig.\ref{ratio}. \\
3. if electron beam current is less than starting value $j<j_{start}$ then
energy in electromagnetic wave at the system entrance can be written in the form:
\begin{figure}[h]
\epsfxsize = 14 cm
\centerline{\epsfbox{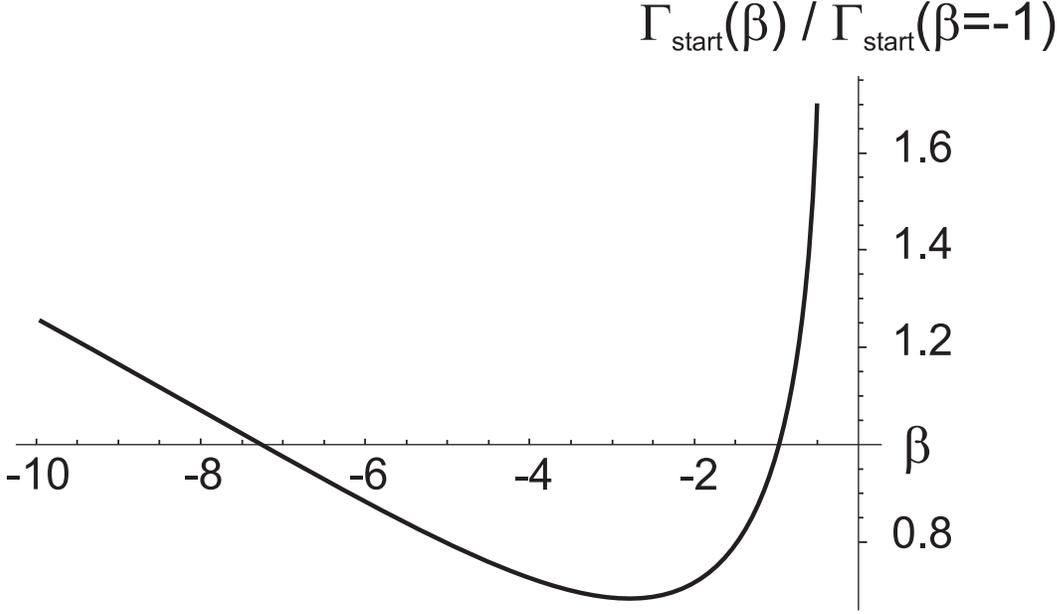}}
\caption{Dependence of form-factor of quasi-Cherenkov superradiation on
asymmetry factor $\beta$.}
\label{ratio}
\end{figure}
\begin{eqnarray}
I/(\gamma _0 |a|^2+|\gamma _1| |b|^2)=1
-\beta \frac{\pi ^{2}n^{2}}{4}\frac{\pi \Theta _{s}^{2}j_{0}c^{2}}{\gamma
_{z}^{2}\gamma I_{A}\omega ^{2}}(kL)^3
\left( \frac{k\chi _{\tau }L}{4\pi }\right) ^{2} f(y)\label{ampl}
\end{eqnarray}
The conventional FEL gain is proportional to
$(kL)^3$ \cite{marsh}, but as follows from (\ref{ampl}) in case of two-wave diffraction
the gain gets an additional factor
$\sim \left( \frac{k\chi _{\tau }L}{4\pi }\right) ^{2}$,
which noticeably exceeds the unity in conditions of dynamical diffraction.
Such increase of radiation output in degeneration point can be explained by the
reduction of wave group velocity, which can be written as:
\begin{eqnarray}
v_{g}=-\left( \frac{\partial D}{\partial k_{z}}\right) /\left( \frac{
\partial D}{\partial \omega }\right)\sim \prod\limits_{i<j}(k_{zi}-k_{zj})  \label{group}
\end{eqnarray}
It follows from (\ref{group}) that for multi-wave dynamical diffraction in the
$s$-fold-degeneration point  $v_g\sim v_0/(kL)^s$, the starting current
$j_{start}\sim (kL)^{-3}(k\chi _1L)^{-2s}$ and {amplification}
is proportional to  $(kL)^3(k\chi _1L)^{2s}$. 
It should be noted that  considered effects take place in wide spectral range 
for wavelengths from centimeters to X-ray (\cite{bar1,bar2,ba2,bar4,Batr})  
and influence of effect increases with the frequency growth.

\section{Conclusion}
The generation threshold in undulator VFEL in case of VDFB  can be achived at 
lower electron beam current and shorter undullator length when special 
conditions of roots degeneration are fulfilled. Change of VDFB conditions by varying 
the volume geometry parameters (for example, the angle between wave vectors) gives the
possibility to increase Q-factor and decrease starting current (see fig. \ref{ratio})
and, hence, the efficiency of generation can be increased.

\end{document}